\documentclass[prb,floats,aps,amssymb,preprint,epsf,epsfig,rotate,superscriptaddress]{revtex4}
\usepackage{bm}
\usepackage{graphicx}
\usepackage{xcolor}

\begin{document}
\title{Asymptotics of the metal-surface Kohn-Sham exact exchange potential revisited}

\author{C. M. Horowitz}
\affiliation{Instituto de Investigaciones Fisicoqu\'imicas Te\'oricas 
y Aplicadas, (INIFTA), UNLP, CCT La Plata-CONICET, Sucursal 4, Casilla de Correo 16, (1900)
La Plata. Argentina}
\author{C. R. Proetto}
\affiliation{Centro At\'omico Bariloche and Instituto Balseiro, 8400, S. C. de Bariloche, R\'{i}o Negro, Argentina}
\author{J. M. Pitarke}
\affiliation{CIC nanoGUNE BRTA, Tolosa Hiribidea 76, E-20018 Donostia, Basque Country, Spain}
\affiliation{Materia Kondentsatuaren Fisika Saila, Centro F\'isica Materiales CSIC-UPV/EHU, and DIPC, 644 Posta Kutxatila, E-48080 Bilbo, Basque Country, Spain}

\date\today

\begin{abstract}
The asymptotics of the Kohn-Sham (KS) exact exchange potential $V_x(z)$ of a jelliumlike semi-infinite metal is investigated, in the framework of the optimized-effective-potential formalism of density-functional theory. Our numerical calculations clearly show that deep into the vacuum side of the surface
$V_x(z) \propto e^2 \ln(az) / z$, with $a$ being a system-dependent constant, thus confirming the analytical calculations reported in
Phys. Rev. B {\bf 81}, 121106(R) (2010). A criticism of this work published in Phys. Rev. B {\bf 85}, 115124 (2012) is shown to be incorrect. Our rigorous exchange-only results provide strong constraints both for the building of approximate exchange functionals and for the determination of the still unknown KS correlation potential.
\end{abstract}


\maketitle

\section{Introduction}

The asymptotics of the Kohn-Sham (KS) exchange-correlation ($xc$) potential $V_{xc}(z)$ of density-functional theory (DFT) at metal surfaces remains an important and open field of research. In their seminal DFT investigation of the electronic structure of metal surfaces, Lang and Kohn~\cite{LK70} pointed out that far outside the surface $V_{xc}(z)$ should behave, such as the classical image potential $- \; e^2/4z$, with $z$ being the distance from the metal-vacuum surface located at $z=0$. This has been widely accepted; but more than 50 yr later there is no rigorous proof of its validity yet. As a first step towards this goal, here we split $V_{xc}(z)$ into its exchange ($x$) and correlation ($c$) contributions, i.e.,
$V_{xc}(z) = V_{x}(z) + V_{c}(z)$, and we focus on a rigorous evaluation of $V_{x}(z)$ in the framework of the so-called optimized-effective-potential (OEP) formalism~\cite{Grabo00,KK08}. These rigorous $x$-only results should then serve as a basis for the determination of the still {\it unknown} KS correlation potential. Besides, and from a more general perspective, exact results are always most welcome in DFT, since a success strategy for the systematic building of progressively more predictive $xc$ energy functionals is based on the satisfaction of as many exact features as possible~\cite{P05,PRSB14,PSMD16}. 

Along these lines, the asympotics of $V_{x}(z)$ have been analyzed in the case of {\it (i)} metal slabs –with a discrete energy spectrum~\cite{HPR06,HPP08,HCPP09,HPP10}-,
and {\it (ii)} an {\it extended} semi-infinite (SI) metal – with a continuous energy spectrum~\cite{HCPP09,HPP10}–. In the vacuum region of a metal slab of width $d$, one finds the following universal asymptotics for the KS exchange potential~\cite{HPR06}:
\begin{equation}
    V_{x}^{\text{Slab}}(z/d \gg 1) \rightarrow - \frac{e^2}{z} \;
    \label{slab}
\end{equation}
and the exchange energy per particle~\cite{HCPP09}:
\begin{equation}
    \varepsilon_{x}^{\text{Slab}}(z/d \gg 1) \rightarrow - \frac{e^2}{2z} \; ,
    \label{slabe}
\end{equation}
which do not depend on the metal electron density. This {\it exact} result is in sharp contrast with the much faster exponential decay displayed by $V_{x}(z)$
in all local or semi-local approximations, such as local density approximation (LDA), generalized gradient approximation (GGA), and meta-GGA~\cite{ed11}, which precludes an accurate evaluation of the electronic structure of metal surfaces~\cite{NP11}. Equations.~(\ref{slab}) and (\ref{slabe}), which were obtained for jellium slabs, have been validated recently for {\it real} slabs in a series of works by Engel~\cite{Engel14a,Engel14b,Engel18a,Engel18b,Engel18c}. In these publications, the jellium approximation was replaced by the real atomic configuration of the slabs by using a plane-wave pseudopotential approach. After averaging over the plane parallel to the surface, the universal Eqs.~(\ref{slab}) and (\ref{slabe}) were found again as in the case of jellium slabs. These rigorous slab asymptotics are very much in line with the corresponding asymptotics for finite systems, such as atoms, molecules, and metal
clusters, where $V_{x}(r) \sim - \; e^2/r$ and
$\varepsilon_{x}(r) \sim - \; e^2/(2r)$ far within the vacuum. In the case of finite systems, correlation is known not to contribute 
to the leading asymptotics, so far enough in the vacuum one can also write
$V_{xc}(r) \sim - \; e^2/r$ and
$\varepsilon_{xc}(r) \sim - \; e^2/(2r)$.~\cite{AvB85,RDGG04,Kraisler20}
It should be noted here that all these rigorous asymptotics are the result of the fact that far enough from the center of the finite system or slab the electron density is dominated by the highest occupied KS orbital. This is not so for the SI metal, as in this case the energy spectrum is continuous and one cannot, therefore, isolate one single KS orbital at the Fermi level. Hence, it is important, in the case of the more subtle SI metal, to combine analytical derivations with fully numerical calculations. This is precisely what we report here, by focusing on the SI metal, as the slab exchange results [Eqs. (\ref{slab}) and (\ref{slabe})] are known to be final and free from any controversy.

In this paper, we revisit the analytical calculations reported in Ref.~[\onlinecite{HPP10}], which we combine here with fully numerical OEP calculations to reach the conclusion that the KS exchange potential far outside a SI metal surface scales indeed as
\begin{equation}
 V_x(z/\lambda_F \gg 1)\to {e^2\over z}\,\ln{(a z)}\; , 
\label{asint.tot}
\end{equation}
where $\lambda_F$ is the Fermi wavelength and $a$ represents a coefficient that depends on the average electron density of the metal. This is in contrast with the analytical OEP derivations reported in Ref.~[\onlinecite{Qian12}] where it is wrongly concluded that the KS exchange potential $V_{x}(z)$ coincides far away from the surface with the exchange energy per particle
$\varepsilon_x(z)$, which is known to be of the image-like form $\varepsilon_{x}(z) \sim - \; e^2 R_x/(2z)$, with $R_x$ being a dimensionless metal-dependent constant to be defined below. 

The paper is organized as follows: In Sec. II, our SI jellium model is introduced, together with the basics
of the OEP formalism, and a few known asymptotics are revisited. Fully numerical calculations of $V_x(z)$ and its various contributions are given in Sec. III, with a particular emphasis on the difficult large-$z$ limit.
In Sec. IV, our main results are summarized.

\section{Model and basics of the OEP method}

We consider a SI jellium model of a metal surface, where the discrete character of the positive ions inside the metal is replaced by a SI uniform positive neutralizing background (the jellium). This positive jellium density is given by 
\begin{equation}
n_{+}(z) = \overline{n} \,\, \theta (-z),
\label{jellium}
\end{equation}
which describes a sharp jellium $(z<0)$ - vacuum $(z>0)$ interface at $z=0$. $\overline{n}$ is a constant with the dimensions of a three-dimensional (3D) density that through the global neutrality condition fixes the electron density.
The model is invariant under translations in the $x-y$ plane (of normalization area $A$), whose immediate consequence is that the 3D KS orbitals of DFT can be factorized as 
\begin{equation}
 \varphi_{{\bf k}_{\parallel},k}({\bf r})=\frac{e^{i{{\bf k}_{\parallel} \cdot} \bm{ \rho }}}{\sqrt{A}}\,
 \frac{\xi_{k}(z)}{\sqrt{L}},
\label{KSfunctions}
\end{equation}
where $\bm{\rho}$ and ${\bf k}_{\parallel}$ are the in-plane coordinate and
wave vector, respectively, and $L$ represents a normalization length along the $z$ direction.
The limit $L \rightarrow \infty$ will be taken below, both for the analytical derivations and for the numerical calculations.
The spin-degenerate orbitals $\xi _{k}(z)$ are the normalized solutions of the effective one-dimensional KS equation:
\begin{equation}
 \widehat{h}_{\text{KS}}^{k}(z) \xi _{k}(z) = 
 \left[-\frac{\hbar ^{2}}{2m}
 \frac{\partial ^{2}}{\partial z^{2}}+\overline{V}_{\text H}(z) + V_{\text {xc}}(z) - \varepsilon_k \right] \xi _{k}(z) = 0,
 \label{KSequations}
\end{equation}
where $\varepsilon_k$ are the KS energies, 
and $\overline{V}_{\text H}(z)$ is the electrostatic {\em effective} Hartree potential~\cite{note1} 
\begin{eqnarray}
 \overline{V}_{\text{H}}(z) &:=& V_{\text{ext}}(z) + V_{\text{H}}(z) \nonumber \; , \\ 
 &=& 2 \pi e^{2}\int_{-L/2}^{L/2}dz^{\prime } \left[ n(z^{\prime })-n_{+}(z^{\prime })\right]
 \int_0^{\infty} \frac{{\rho}^{\prime} d{\rho}^{\prime}}{\sqrt{(\rho - \rho^{\prime})^2 + (z-z^{\prime})^2}} \; ,\nonumber \\
 &=& -2 \pi e^{2}\int_{-L/2}^{L/2}dz^{\prime }
 \left|z-z^{\prime }\right| \left[ n(z^{\prime })-n_{+}(z^{\prime })\right] \;,
 \label{hartree}
\end{eqnarray}
with $n(z)$ being the ground-state electron density
\begin{equation}
 n(z) = 2 \sum_{{\bf k}_{\parallel},k}^{occ} |\varphi_{{\bf k}_{\parallel},k}({\bf r})|^2 = 
 \frac{1}{4 \pi^2} \int_{-k_F}^{k_F} (k_F^2-k^2) |\xi_{k}(z)|^2 dk \; .
 \label{density}
\end{equation}
The Fermi wave number $k_F$ is determined by imposing the "bulk" neutrality condition that the integral of $n(z)$ along $z$
($ |z| \leq L/2$) be equal to $\overline{n} L$; thus, $k_F = (3 \pi^2 \overline{n})^{1/3}$. The
Fermi wavelength is simply $\lambda_F=2\pi/k_F$.
Note that (i) in passing from the second to the third line in Eq.({\ref{hartree}}) the bulk neutrality condition has been used to
nullify a divergent contribution arising from the $\rho^{\prime}$ integration and (ii) in Eq.(\ref{density}) the discrete sums over
${\bf k}_{\parallel}$ and $k$ have been converted into the usual integrals using the conversion factors $A/(2\pi)^2$ and $L/(2\pi)$,
respectively.  

$V_{xc}(z)$ is the so-called $xc$ potential, which is obtained as the functional derivative of the $xc$-energy functional $E_{xc}[n(z)]$; 
for our {\em effective one-dimensional} semi-infinite system :~\cite{fd} 
\begin{equation}
 V_{xc}(z) = \frac{1}{A} \; \frac{\delta E_{xc}}{\delta n(z)} \; .
 \label{fd}
\end{equation}
Within the OEP framework, $V_{xc}(z)$ may be expressed as follows:
\begin{equation}
 V_{xc}(z) = V_{xc}^{\text{KLI}}(z) + V_{xc}^{\text{Shift}}(z) \; ,
 \label{splitting}
\end{equation}
where $V_{xc}^{\text{KLI}}(z)$ represents the so-called Krieger-Li-Iafrate (KLI) contribution~\cite{KLI},
\begin{equation}
 V_{xc}^{\text{KLI}}(z) = \int_0^{k_F} \frac{|\xi_{k}|^2}{2 \pi^2 n(z)}
 \left[ u_{xc}^{k}(z) + \overline{\Delta V}_{xc}^{k} \right] \widetilde{dk} \; ,
 \label{KLI}
\end{equation}
with $\widetilde{dk} = (k_F^2 - k^2) dk$, $\Delta V_{xc}^{k}(z) = V_{xc}(z)-u_{xc}^k(z)$, and $u_{xc}^k(z) = 
[4\pi/A(k_F^2-k^2)\xi_k^*(z)]\delta E_{xc} / \delta \xi_k(z)$; the $u_{xc}^k(z)$'s are usually referred to as orbital-dependent $xc$ potentials. 
Besides~\cite{HPP10},
\begin{equation}
V_{xc}^{\text{Shift}}(z) = - \int_0^{k_F}
\frac{\left[ (k_F^2-k^2)\Psi_k(z)\xi_k(z)+\Psi'_k(z)\xi'_k(z)\right]}{2\pi^2n(z)}\widetilde{dk} \; , 
\label{shift}
\end{equation}
with primes denoting derivatives with respect to the coordinate $z$, and
$\Psi_k(z)$ are the so-called orbital shifts; they are defined as the solutions of the following differential equation~\cite{note4}:
\begin{equation}
 \widehat{h}_{\text{KS}}^k(z)\Psi_k(z) = - \left[\Delta V_{xc}^k(z) - \overline{\Delta V}_{xc}^k  \right] \xi_k(z) \; ,
 \label{de}
\end{equation}
or, equivalently, by
\begin{equation}
\Psi_k(z) = \int_{-L/2}^{L/2} \xi_k(z') \Delta V_{xc}^k(z')G_k(z',z)~dz'\; ,
\label{shift1}
\end{equation}
with $G_k(z,z')$ being the KS Green function~\cite{note5}:
\begin{equation}
G_k(z,z') = \frac{1}{\pi} P \int_{0}^{\infty} \frac{\xi_{k'}^*(z)\xi_{k'}(z')}{(\varepsilon_k-\varepsilon_{k'})}~dk'.
\label{Green}
\end{equation} 

Equations~(\ref{KSequations}) - (\ref{de}) form a closed set of equations, whose self-consistent solutions must be obtained numerically, once the
key $xc$-energy functional $E_{xc}[n]$ is provided.  Importantly, this set of equations determines $V_{xc}(z)$ up to an
additive constant that should be fixed by imposing some suitable boundary condition;  this fact is a consequence of the
metal surface being represented as a closed system~\cite{RHP15}.
Since $E_{xc}[n] = E_{x}[n] + E_{c}[n]$, which in turn implies that 
$V_{xc}(z) = V_x(z) + V_c(z)$, we will now focus on the exchange contribution $V_x(z)$ to the KS $xc$ potential of Eq.~(\ref{splitting}),
leaving the more complicated correlation contribution for further studies. The main reason for this procedure is that since
$E_x[n]$ is known {\it exactly} in terms of the KS orbitals $\xi_k(z)$, it is feasible to obtain the {\it exact} $V_x(z)$ as the self-consistent solution of the exchange-only version of Eqs.~(\ref{KSequations}) - (\ref{de}).

In the case of a SI system,
\begin{equation}
 E_x[n] = A \int_{-L/2}^{L/2} dz \, n(z) \, \varepsilon_x(z) \; , 
 \label{ex}
\end{equation}
where $\varepsilon_x(z)$ is the position-dependent exchange energy per particle at plane $z$. This quantity, which has been studied in detail in Ref.~[\onlinecite{HCPP09}], is given by
\begin{equation}
\varepsilon_{x}(z) = -\frac{e^2}{2 \pi^2 n(z)}
\int\limits_{-k_{F}}^{k_{F}} dk \int\limits_{-k_{F} }^{k_{F}} dk^{'} (k_{F}^{2}-k^{2})^{1/2}
(k_{F}^{2}-k^{'2})^{1/2} \int\limits_{-L/2}^{L/2}
dz^{\prime }\gamma_{k}(z,z^{\prime })\gamma _{k^{'}}(z^{\prime
},z)F_{k,k^{'}}(z,z^{\prime }),
\label{5.semi}
\end{equation}
where $\gamma_{k}(z,z^{\prime })=\xi _{k}(z)^{*}\xi_{k}(z^{\prime })$
and $F_{k,k^{'}}(z,z^{\prime })$ is the Kohn-Mattsson function~\cite{KM98}, whose explicit expression is 
\begin{equation}
 F_{k,k'}(z,z') = \frac{1}{4 \pi} \int_0^{\infty} \frac{d\rho}{\rho} 
 \frac{J_1\left[\rho(k_F^2-k^2)^{1/2}\right] J_1\left[\rho(k_F^2-{k'}^2)^{1/2}\right]}{[\rho^2 + (z-z')^2]^{1/2}} \; ,
\end{equation}
with $J_1(x)$ being the cylindrical Bessel function of the first kind.
In most DFT standard approximations, $\varepsilon_x(z)$, which is a functional of the electron density $n(z)$, is obtained from the knowledge of the exchange energy per particle of a uniform electron gas at $z$ and nearby, thus leading to local (LDA) or semi-local (GGA) approximations of $\varepsilon_x(z)$ and $V_x(z)$ that decay exponentially far into the vacuum side of the surface and fail badly to describe the actual asymptotics of these quantities which are strongly non-local in nature.

The exact KS exchange potential is obtained [see Eq.~(\ref{fd})] as the functional derivative of the exact exchange energy of Eq.~(\ref{ex}). As $\varepsilon_x(z)$ of Eq.~(\ref{5.semi}) depends explicitly on the KS orbitals but only implicitly on the electron density $n(z)$, we follow the OEP method, which allows us to deal with orbital-dependent energy functionals. Indeed, the OEP method allows us to obtain the exact KS exchange potential and, therefore, its actual asymptotics far into the vacuum side of the surface. 

In order to make contact with previous discussions, we write
\begin{equation}
 V_x(z) = V_x^{\text{KLI}}(z)+V_x^{\text{Shift}}(z) \; ,
\label{vexchange}
\end{equation}
and we split the exchange part of the KLI $xc$ potential of Eq.~(\ref{KLI}) as follows: 
\begin{equation}
 V_x^{\text{KLI}}(z) = V_x^{\text S}(z) + V_x^{\Delta}(z) \; ,
\label{KLI0}
\end{equation}
where $V_x^{\text S}(z) = 2\,\epsilon_x(z)$ is the so-called Slater potential involving the orbital-dependent exchange potentials $u_x^k(z)$,
whereas $V_x^{\Delta}(z)$ represents the contribution from $\overline{\Delta V}_x^{\, k}$.
The corresponding asymptotics are~\cite{SS97,Nastos,QS05} as follows:
\begin{equation}
 V_x^{\text S}(z/\lambda_F \gg 1) \rightarrow - \; 
 \left(\frac{\pi+2 \alpha \ln{\alpha}}{\pi(1+\alpha^2)}\right) \frac{e^2}{z}=
 - \; {R_x} \frac{e^2}{z} \; ,
 \label{asint.slater}
\end{equation}
where $\alpha = a_0 k_F/\sqrt{2W}$, $a_0$ and $W$ being the Bohr radius and the metal-vacuum work function —in atomic units—, and~\cite{HPP10}
\begin{equation}
 V_x^{\Delta}(z/\lambda_F \gg 1) \rightarrow \frac{e^2}{2 \pi \alpha z}[\ln(\alpha k_F z) + C] \; , 
  \label{asint.delta}
\end{equation}
where $C \sim 0.963 51$. Equation~(\ref{asint.delta}) is obtained by replacing $\Delta V_x^k(z)$, entering the calculation of $\overline{\Delta V}_x^k$, by its bulk value [$\Delta V_x^k(z) \simeq - \; e^2 k_F/\pi -u_x^k(z/\lambda_F \rightarrow - \infty)$], which is fully justified, as for a SI metal the mean value is not sensitive to the actual form of the KS orbitals near the metal surface and far into the vacuum. An explicit analytical expression for
$u_x^k(z / \lambda_F \rightarrow - \infty)$ is obtained 
through a  ${\bf k}_\parallel$ Fourier transform 
of the orbital-dependent exchange potential $u_x$ of a 3D electron gas~\cite{HPP10,bardeen}.
The all-important boundary condition commented above is satisfied by imposing this choice for the bulk value of the exchange potential in the OEP numerical procedure. By doing so, the related additive constant to $V_x(z)$ is fixed in such a way that  $V_{x}(z/\lambda_F \gg 1) \rightarrow 0$.
Below, we show how these analytical asymptotics [Eqs.~(\ref{asint.slater}) and (\ref{asint.delta})] are
accurately confirmed by our numerical OEP calculations.

Our numerical strategy to solve the SI metal-surface OEP equations starts with the definition of the following three regions: far-left bulk region ($-L/2 < z < z_1$, with $L \to \infty$), central
region (a few $\lambda_F$'s to
the left and to the right of the jellium edge), and far-right vacuum region ($z_N < z < L/2$). In the bulk
region, the KS eigenfunctions are taken to be of the form $\xi _{k}(z) =\sin(kz-\delta_{k})$,
where $\delta_{k}$ are phase shifts, thus fixing an overall normalization constant~\cite{LK70}.
By doing that, the system becomes effectively infinite along the negative-$z$ direction.
In the central region, we define a mesh of $N$ points between
$z_1$ and $z_N$ $(z_1 < z_N )$, the first point $z_1$ being chosen
far enough from the jellium edge, in the bulk, so that the Friedel
oscillations can be neglected, and the outer point $z_N$ being chosen to be far
enough from the jellium edge into the vacuum so that the effective one-electron
potential is negligibly small. 
Since the KS potential $\overline V_{\text{H}}(z)+V_{xc}(z) \sim 0$ for $z \geq z_N$, the
KS eigenfunctions can be approximated as $\xi_{k}(z) = s \, e^{-k^{*}z}$, where $s$ is a
constant and
$k^{*} = (-2 m \varepsilon_{k} / \hbar^2)^{1/2}$. At the
mesh points, the KS orbitals $\xi_{k}(z)$ are calculated by using the Numerov integration procedure~\cite{liebsch}.
By doing that, the system becomes effectively infinite also along the positive-$z$ direction.
Regarding the orbital shifts, they are calculated 
from Eq.~(\ref{shift1}), with the KS Green function being computed by using the procedure described in
Ref.~[\onlinecite{1liebsch}]. For our systems we have found it convenient to place $z_1$ at a distance of 5 $\lambda_F$ to
the left of the jellium edge ($z_1 = -5 \lambda_F$), $z_N$ at a distance of 25 $\lambda_F$ to
the right of the jellium edge ($z_N = 25 \lambda_F$) and a mesh of $1200$ points. The approximation $\xi_{k}(z) = s \, e^{-k^{*}z}$
quoted above is used only for $z > z_N$, but we have checked that it is already valid for smaller values of $z$. 
See Ref.~[\onlinecite{HCPP09}] for more numerical details.

\begin{figure}
 \includegraphics[width=0.75\columnwidth,angle=0]{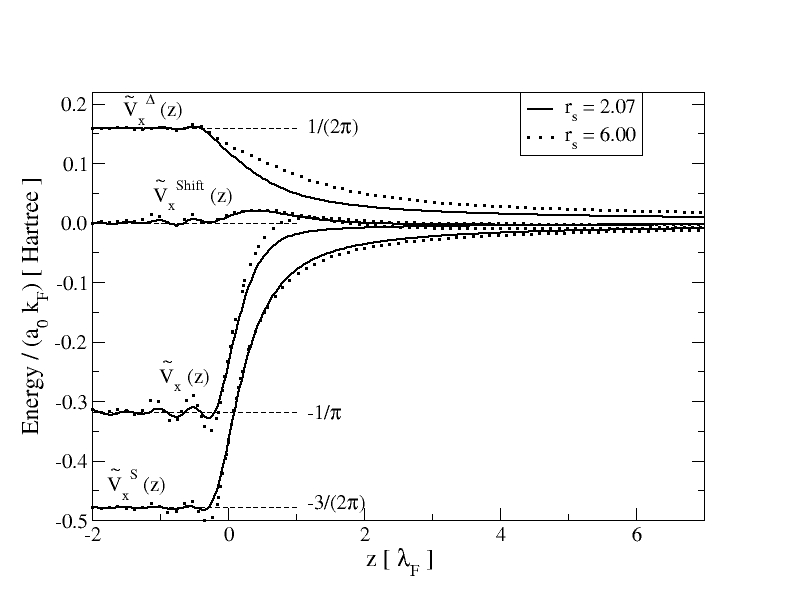}
 \caption{OEP self-consistent calculations of $\tilde V_x(z)=V_x(z)/(a_0k_F)$ and its three contributions $\tilde V_x^{\text{S}}(z)$, $\tilde V_x^{\Delta}(z)$,
 and $\tilde V_x^{\text{Shift}}(z)$,
 for $r_s=2.07$ (solid lines) and $r_s=6.00$ (dotted lines). The jellium-vacuum interface is at $z=0$. The horizontal dashed lines on the vertical axis represent
 the corresponding {\it universal} bulk limits (in Hartree atomic units): $V_x(z/\lambda_F \to-\infty) / (a_0 k_F) = -1/\pi\simeq -0.318$, 
 $V_x^{\text{S}}(z/\lambda_F \to-\infty)/(a_0 k_F) = -3/2\pi\simeq -0.477$~~\cite{bardeen},
 $V_x^{\Delta}(z/\lambda_F \to-\infty)/(a_0 k_F) = 1/2\pi\simeq 0.159$~~\cite{GLLB95}, and
 $V_x^{\text{Shift}}(z/\lambda_F \to -\infty) = 0$~~\cite{note3}. 
 }
 \label{fig.1}
\end{figure}

As an example of our numerical OEP calculations, in Fig.~\ref{fig.1} we display $V_x(z)$ and its three contributions:
$V_x^{\text S}(z)$, $V_x^\Delta(z)$, and $V_x^{\rm Shift}(z)$,
for a high-density metal with $r_s = 2.07$ and a low-density metal with
$r_s = 6.00$.~\cite{note2} All contributions have been scaled with the factor $(a_0 k_F)^{-1}$, as the scaled $V_x(z)$ yields then an "universal" (material independent) value in the bulk:
$V_x(z/\lambda_F\to-\infty)/(a_0k_F)=-1/\pi$.
It is interesting to note the presence of marked Friedel-like oscillations in the case of
$r_s = 6.00$, which are known to be due to the poor screening capability  
of the electron gas at this low electron density. Within this context, the "shoulder" exhibited, for $r_s = 6.00$ but not for $r_s = 2.07$, by $V_x(z)$ right after the interface on the vacuum side of the surface may be considered as a last enhanced oscillation that becomes indeed a shoulder.

\section{Numerical asymptotics}

First of all, we proceed with the OEP numerical study of the two contributions entering the KLI exchange potential of
Eq.~(\ref{KLI0}) as this serves as a strong test for our OEP calculations and allows us to estimate the regime where  $V_x(z)$ reaches its asymptotic behavior.

\begin{figure}
 \includegraphics[width=0.75\columnwidth,angle=0]{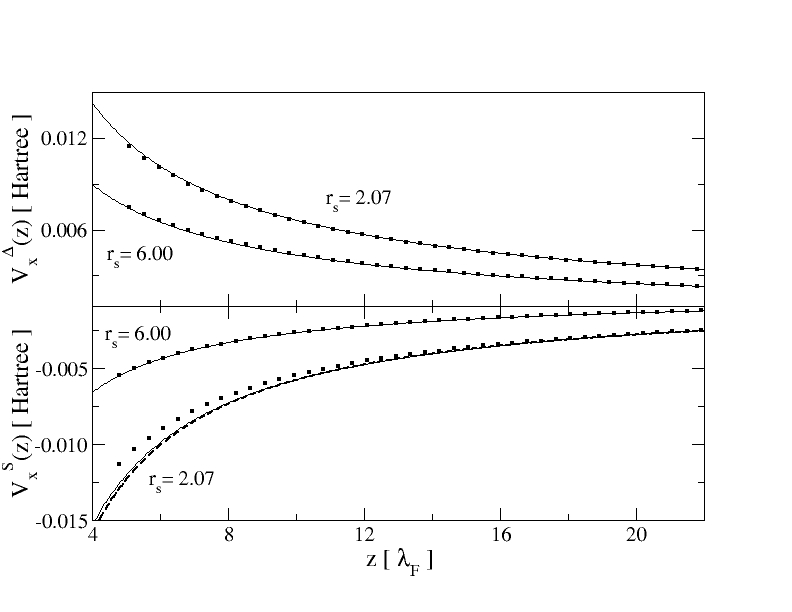}
 \caption{Upper panel: OEP self-consistent calculations of $V_x^{\Delta}(z)$
 (in Hartree atomic units) for $r_s=2.07$ and $r_s=6.00$ (solid lines), and their 
 analytic asymptotics of Eq.~(\ref{asint.delta}) (dotted lines). Lower panel: OEP self-consistent calculations of $V_x^{\text{S}}(z)$ (in Hartree atomic units) for $r_s=2.07$ and $r_s=6.00$ (solid lines), and their analytic asymptotics of Eq.~(\ref{asint.slater}) (dotted lines). The dashed curve for $r_s = 2.07$ represents the sum of the analytic asymptotics of Eq.~(\ref{asint.slater}) and the contribution from the penetration of the
 vacuum orbitals (see the main text) that is neglected in the derivation of Eq.~(\ref{asint.slater}).}
 \label{fig.2}
\end{figure}

\begin{figure}
 \includegraphics[width=0.75\columnwidth,angle=0]{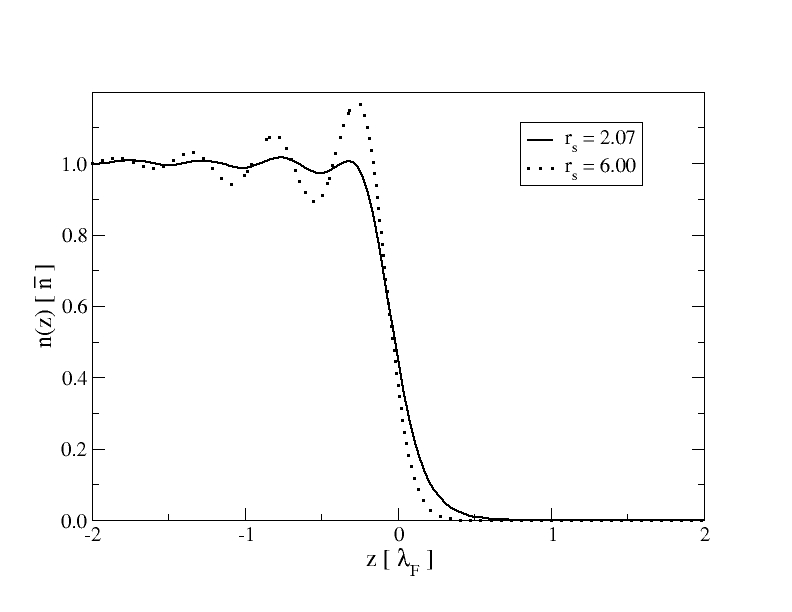}
 \caption{
 Electron density $n(z)$ for $r_s=2.07$ (solid line) and $r_s=6.00$ (dotted line). 
 The jellium edge is located at $z=0$. In both cases, the electron density is normalized with the bulk density $\overline{n}$. By using $\lambda_F$ as the length unit, the comparison (Friedel oscillations and spill out) between both electron densities becomes feasible. 
 }
 \label{fig.3}
\end{figure}

Figure~\ref{fig.2} shows the result of our OEP calculations
of $V_x^{\text{S}}(z)$ and $V_x^{\Delta}(z)$ for $r_s=2.07$ and $r_s=6.00$, together with the corresponding analytical asymptotics of Eqs.~(\ref{asint.slater}) and (\ref{asint.delta}). In the case of $V_x^{\Delta}(z)$, the agreement between our fully numerical calculations and the analytical asymptotics is
excellent for both electron densities at distances from the surface that are beyond 6 $\lambda_F$. In the case of $V_x^{\text{S}}(z)$ and for high electron densities ($r_s=2.07$), however, the analytical asymptotics deviate, even at distances from the surface that are beyond 6 $\lambda_F$, from the actual numerical calculations. A detailed numerical study shows that this slight deviation is a consequence of the extension of the KS orbitals close to the Fermi energy beyond the interior of the metal. More explicitly, to obtain the analytical asymptotics of Eq.~(\ref{asint.slater}), it is assumed that the main contribution in Eq.~(\ref{5.semi})
comes from inside the metal and, accordingly, the vacuum region in the $z^{\prime }$ integral is neglected~\cite{SS97,Nastos,QS05}. 
This approximation starts to fail at high electron densities ($r_s = 2.07$) where the orbitals extend considerably into the vacuum, as in this case the neglected contribution to the integral of Eq.~(\ref{5.semi}) starts to play a role.
It is well known that at high electron densities the electron spill out of the surface is more pronounced, as seen in Fig.~\ref{fig.3}, where the electron-density spill out is
represented  for both $r_s = 2.07$ and $r_s=6.00$. Indeed,
when the numerical contribution from the vacuum region near the surface is added (for $r_s=2.07)$ to Eq.~(\ref{asint.slater}), we find the
dashed curve in the lower panel of Fig.~2, which nicely agrees with the full numerical calculation for all distances beyond 6 $\lambda_F$. 

\begin{figure}
 \includegraphics[width=0.75\columnwidth,angle=0]{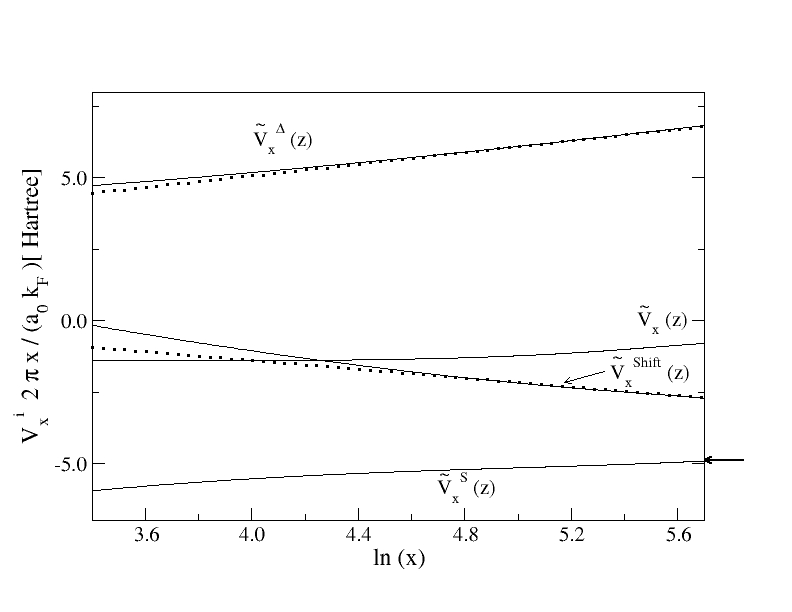}
 \caption{OEP self-consistent calculations of $\tilde V_x^{\Delta}(x)=(2\pi\,x/a_0k_F)\,V_x^{\Delta}(x)$ and
 $\tilde V_x^{\text{Shift}}(x)=(2\pi\,x/a_0k_F)\,V_x^{\text{Shift}}(x)$ for $r_s=2.07$ (solid lines), with
 $x=\alpha k_F z$. The dotted lines represent numerical fits from Eq.~(\ref{fit}) with the fitting parameters given in the text. OEP self-consistent calculations of
 $\tilde V_x^{\text{S}}(x)=(2\pi\,x/a_0k_F)\,V_x^{\text{S}}(x)$ and $\tilde V_x(x)=(2\pi\,x/a_0k_F)\,V_x(x)$ are given for comparison. The arrow
 on the right represents the material-dependent asymptotic constant value of $(2\pi\,x/a_0k_F)\,V_x^{\text{S}}(x \gg 1)$, which for $r_s=2.07$ is: $- 2 \pi \alpha R_x \; [(e^2/a_0)] \simeq - \; 4.85 \; [(e^2/a_0)]$.}
 \label{fig.8}
\end{figure}

\begin{figure}
 \includegraphics[width=0.75\columnwidth,angle=0]{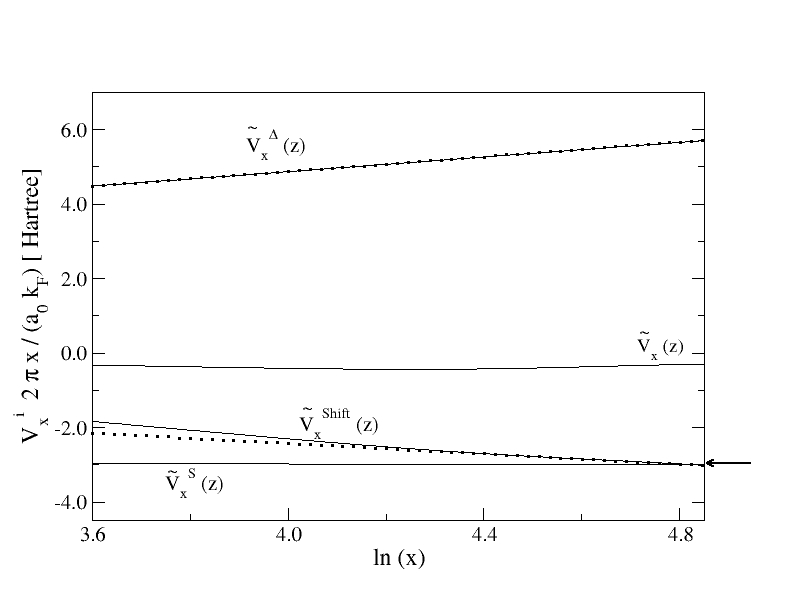}
 \caption{As in Fig.~\ref{fig.8}, but for $r_s = 6.00$. The arrow on the right represents the asymptotic constant value of $(2\pi\,x/a_0k_F)\,V_x^{\text{S}}(x>>1)$, which for $r_s=6.00$ is: $- 2 \pi \alpha R_x \; [(e^2/a_0)] \simeq - \; 2.98 \; [(e^2/a_0)]$. In this case ($r_s=6.00$), our OEP calculations do not go as deep into the vacuum as in the case of $r_s=2.07$
 (Fig.~\ref{fig.8}), which is due to the fact that for $r_s=6.00$ the electron density decays faster outside the metal (see Fig.~\ref{fig.3}) and instabilities associated with the vanishing electron density appear, therefore, earlier in this case.}
 \label{fig.9}
\end{figure}

Now we focus on the {\it last} -and most subtle- exact-exchange contribution $V_x^{\text{Shift}}(z)$ entering Eq.~(\ref{vexchange}) [beyond the KLI contribution of Eq.~(\ref{KLI0})], which
we plot by a solid line in Figs.~\ref{fig.8} and \ref{fig.9}, together with the {\it second}
exact-exchange contribution $V_x^{\Delta}(z)$ entering Eq.~(\ref{KLI0}), well outside the surface up to many Fermi wavelengths, after the scaling $\tilde V^i(x)=(2\pi\,x/a_0k_F)\,V^i(x)$ with
$ x = \alpha k_F z$ . With this scaling, the so-called Slater potential $\tilde V_x^{\text{S}}(x)$ (also plotted in Figs.~\ref{fig.8} and \ref{fig.9}) yields simply
[see Eq.~(\ref{asint.slater})] a material-dependent constant value at a few Fermi wavelengths outside the surface.

Unlike the scaled $\tilde V_x^{\text{S}}(x)$ Slater potential, $\tilde V_x^{\Delta}(x)$ and $\tilde V_x^{\text{Shift}}(x)$ do not yield a constant value in the asymptotic region outside the surface. Instead, they are found to be well fitted, in the asymptotic region, to the following fitting equation:
\begin{equation}
 (e^2/a_0) [a^i + b^i \ln(x)] \; ,
 \label{fit}
\end{equation}
with $a^i$ and $b^i$ being dimensionless fitting parameters ($i=\Delta,\text{Shift}$). In the case of
the {\it second} contribution entering Eq.~(\ref{KLI0}) ($i=\Delta$), one finds $a^{\Delta} = 0.94 \; (\pm \; 0.03)$ and $b^{\Delta} = 1.02 \; (\pm \; 0.02)$ for $r_s=2.07$, and $a^{\Delta} = 0.96 \; (\pm \; 0.03)$ and $b^{\Delta} = 0.99 \; (\pm \; 0.02)$ for $r_s=6.00$, which perfectly agrees, within error bars, with the universal coefficients
$a_{\text exact}^{\Delta}=0.96351$ and $b_{\text exact}^{\Delta}=1$ entering Eq.~(\ref{asint.delta}). In the case of the {\it last} contribution to Eq.~(\ref{vexchange}) ($i=\text{Shift}$), one
finds $a^{\text{Shift}} = 1.70 \; (\pm \; 0.03)$ and $b^{\text{Shift}} = -0.77 \; (\pm \; 0.02)$ 
for $r_s=2.07$, and $a^{\text{Shift}} = 0.34 \; (\pm \; 0.05)$ and 
$b^{\text{Shift}} = -0.70 \; (\pm \; 0.03)$ for $r_s=6.00$. Figs.~\ref{fig.8} and \ref{fig.9}
clearly show that the numerical fit is essentially not distinguishable from the actual OEP calculations for $x>100$  and $x>70$ for $r_s=2.07$ and $r_s=6.00$, respectively. For the electron densities under study ($r_s=2.07$ and $r_s=6.00$), the full coefficient ($b^{\Delta}+b^{\text{Shift}}$) is positive, so we conclude that for these densities
the leading asymptotic contribution to $V_x(z)$ is indeed of the form given
by Eq.~(\ref{asint.tot}), as anticipated in Ref.~[\onlinecite{HPP10}] and in contrast with the main result of Ref.~[\onlinecite{Qian12}].

In Ref.~[\onlinecite{Qian12}], by taking only the contribution to the KS exchange potential from the KS eigenfunction at $k=k_F$ and after a number of approximations, Qian concluded that in the case of a SI metal surface the asymptotics of $V_x(z)$ are embodied by half the Slater potential 
$V_x^{\text S}(z)$, i.e., $V_x^{\text Q}(z/\lambda_F \gg 1) = V_x^{\text S}(z/\lambda_F \gg 1)/2$ (see Eq.~(1) of Ref.~[\onlinecite{Qian12}]). Qian also reproduced Eq.~(\ref{asint.delta}) above (originally reported in Ref.~[\onlinecite{HPP10}]); but
he stated (without proof, see Eq.~(88) of Ref.~[\onlinecite{Qian12}]) that
\begin{equation}
 V_x^{\text{Shift,Q}}(z/\lambda_F \gg 1) = -\frac{1}{2} \; V_x^{\text S}(z/\lambda_F \gg 1) - 
 V_x^{\Delta}(z/\lambda_F \gg 1) \; ,
 \label{claim}
\end{equation}
in such a way that [see Eqs.~(\ref{vexchange}) and (\ref{KLI0}) above]
\begin{equation}
V_x^{\text{Q}}(z/\lambda_F \gg 1)=\frac{1}{2} \; V_x^{\text S}(z/\lambda_F \gg 1) \; .
\end{equation}

\begin{figure}
 \includegraphics[width=0.75\columnwidth,angle=0]{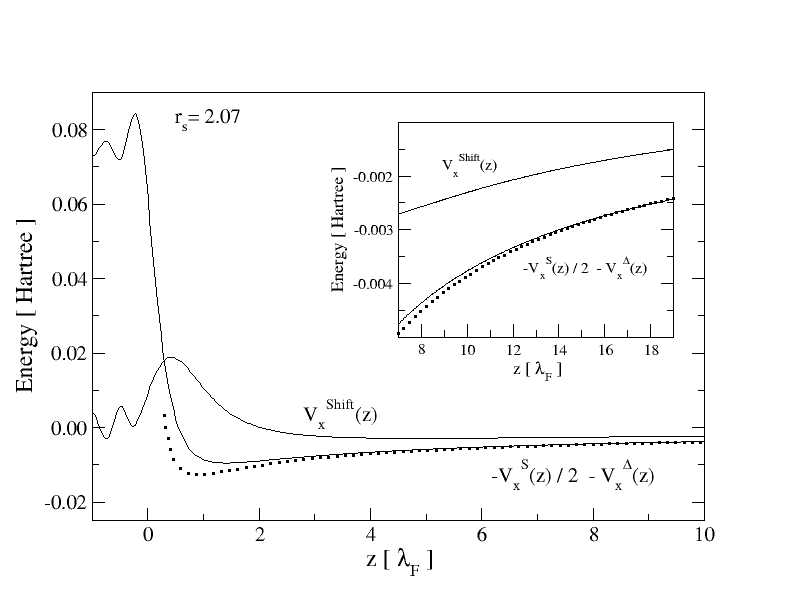}
 \caption{OEP self-consistent calculations of $V_x^{\text{Shift}}(z)$ and $V_x^{\text{Shift,Q}}(z)$, for $r_s=2.07$, the latter computed as in Eq.~(\ref{claim}), i.e.,
 $V_x^{\text{Shift,Q}}(z)=-V_x^{\text S}(z)/2 - V_x^{\Delta}(z)$. The dotted line represents the analytical asymptotics
 of $V_x^{\text{Shift,Q}}(z)$, as obtained by introducing Eqs.~(\ref{asint.slater}) and (\ref{asint.delta}) into Eq.~(\ref{claim}). The inset displays an enlarged view of the far-vacuum region.}
 \label{fig.4}
\end{figure}

\begin{figure}
 \includegraphics[width=0.75\columnwidth,angle=0]{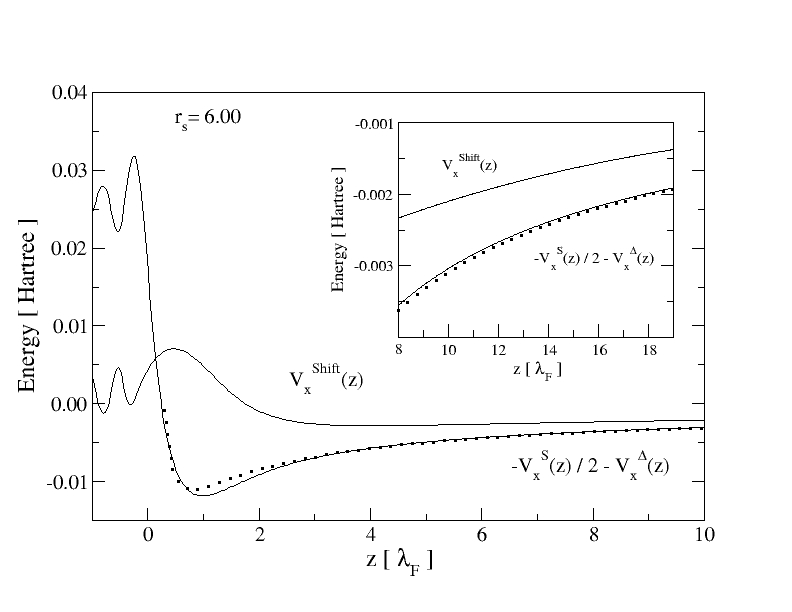}
 \caption{As in Fig.~\ref{fig.4}, but for $r_s = 6.00$.}
 \label{fig.5}
\end{figure}

Figs.~\ref{fig.4} and \ref{fig.5} clearly show that
the actual $V_x^{\text{Shift}}(z/\lambda_F \gg 1)$ and Qian's version
$V_x^{\text{Shift,Q}}(z/\lambda_F \gg 1)$ given by Eq.~(\ref{claim}) do not coincide, which means that
Eqs.~ (1) and (88) of Ref.~[\onlinecite{Qian12}] are simply wrong. In these figures, we plot (for $r_s=2.07$ and $r_s=6.00$) our numerical OEP calculation of the actual $V_x^{\text{Shift}}(z)$ together with our numerical OEP calculation of
Qian's quantity $V_x^{\text{Shift,Q}}(z)=- V_x^{\text S}(z)/2 - V_x^{\Delta}(z)$. From these plots,
it is clear that $V_x^{\text{Shift}}(z)$ and $V_x^{\text{Shift,Q}}(z)$ are two different functions in the asymptotic region. We emphasize here that $V_x^{\text {Shift,Q}}(z)$ of Eq.~(\ref{claim})
has been plotted by taking the numerically exact $V_x^{\text S}(z)$ and $V_x^{\Delta}(z)$
functions, as obtained with our OEP code for all possible values of $z$,
from deep into the bulk ($z/\lambda_F \rightarrow -\infty$) to far into the vacuum ($z/\lambda_F \rightarrow \infty$). Also plotted in these figures (by a dotted line) is the corresponding
$V_x^{\text {Shift,Q}}(z/\lambda_F \gg 1)$ that one obtains by introducing Eqs.~(\ref{asint.slater}) and (\ref{asint.delta}) into Eq.~(\ref{claim}). The asymptotic limit of $V_x^{\text {Shift,Q}}(z)$ is nicely reached at $z/\lambda_F \sim 2$; but it does not coincide with the actual
$V_x^{\text {Shift}}(z/\lambda_F \gg 1 )$

\begin{figure}
\includegraphics[width=0.75\columnwidth,angle=0]{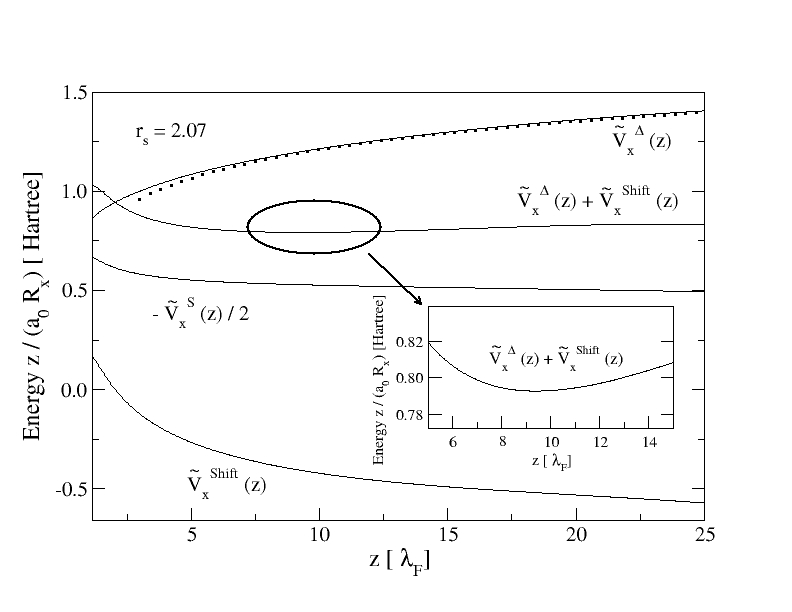}
\caption{OEP self-consistent calculations of the scaled quantities
$\tilde{V}_x^{\Delta}(z) + \tilde{V}_x^{\text {Shift}}(z)$
= $[V_x^{\Delta}(z) + V_x^{\text {Shift}}(z)](z/a_0R_x)$ and
$- \; \tilde{V}_x^{\text S}(z)$ =
$- \; V_x^{\text S}(z)(z/a_0R_x)/2$, for $r_s=2.07$ (solid lines).
Separate OEP self-consistent calculations of the scaled quantities
$V_x^{\Delta}(z)(z/a_0R_x)$ and $V_x^{\text S}(z)(z/a_0R_x)$ are also plotted for comparison.
The dotted line represents the analytical asymptotics of
$V_x^{\Delta}(z)(z/a_0R_x)$, as obtained from
Eq.~(\ref{asint.delta}). The inset displays an enlarged view of the minimum in the curve $[V_x^{\Delta}(z)+V_x^{\text{Shift}}(z)](z/a_0R_x)$.}
 \label{fig.6}
\end{figure}

\begin{figure}
 \includegraphics[width=0.75\columnwidth,angle=0]{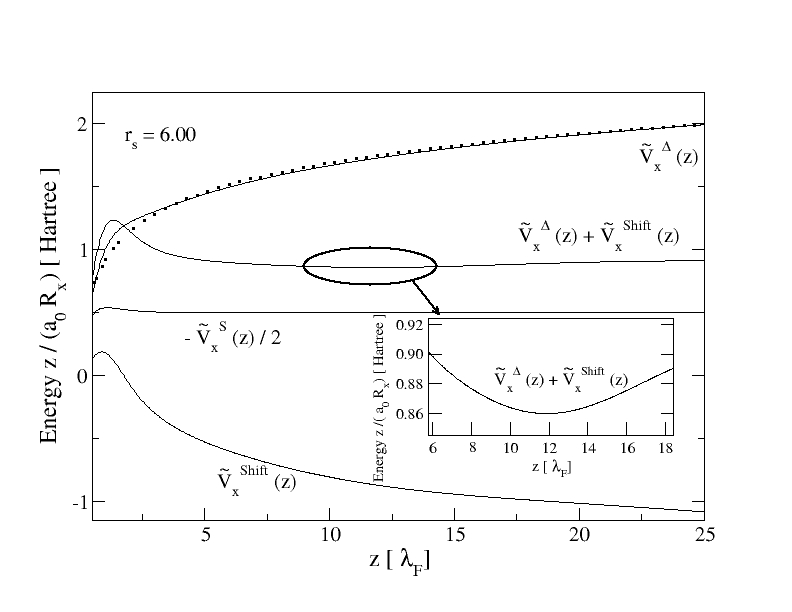}
 \caption{As in Fig. 8, but for $r_s = 6.00$.}
 \label{fig.7}
\end{figure}

Figs.~\ref{fig.6} and \ref{fig.7} show additional evidence of the fact that
Eqs.~(1) and (88) of Ref.~[\onlinecite{Qian12}] are not correct. According to Eq.~(\ref{claim}) above (Eq.~(88) of Ref.~[\onlinecite{Qian12}]):
\begin{equation}
V_x^{\Delta}(z/\lambda_F \gg 1) + V_x^{\text {Shift,Q}}(z/\lambda_F \gg 1)\rightarrow -V_x^{\text S}(z/\lambda_F \gg 1) / 2 = (R_x/2)e^2 / z.
\label{new}
\end{equation}
Hence, if the actual $V_x^{\text{Shift}}(z)$ were to coincide
{\it far outside the surface} with $V_x^{\text{Shift,Q}}(z)$, as claimed by Qian in
Ref.~[\onlinecite{Qian12}]), the scaled quantity $[V_x^{\Delta}(z) + V_x^{\text {Shift}}(z)](z/a_0R_x)$ [see left-hand side of Eq.~(\ref{new})] should approach the constant value 1/2 (in atomic units) as $z/\lambda_F \gg 1$. Figs.~\ref{fig.6} and \ref{fig.7} clearly show that this is not the case. Indeed, the scaled quantity
$[V_x^{\Delta}(z) + V_x^{\text {Shift}}(z)](z/a_0R_x)$ exhibits a minimum 
at $z / \lambda_F \sim 9$ and $z / \lambda_F \sim 12 $, for $r_s = 2.07$ and $r_s=6.00$, respectively. As the scaled $V_x^{\Delta}(z)$ has a positive slope in the displayed
range of $z$, while the opposite is true for the scaled $V_x^{\text {Shift}}(z)$, the existence of a minimum implies that the large-$z$ behavior of
$V_x^{\Delta}(z)$ overcomes the large-$z$ contribution of $V_x^{\text {Shift}}(z)$, and in fact provides the asymptotic limit for
$V_x(z)$, as stated in Ref.~[\onlinecite{HPP10}].

\section{Conclusions}
We have presented a detailed analysis of the KS exchange potential $V_x(z)$ at a jellium-like
SI metal surface, by numerically solving the corresponding OEP equations of DFT. Our numerical calculations clearly show that, at least for the metal electron densities under study, deep into the vacuum side of the surface
$V_x(z) \propto e^2 \ln(az)/z$, for $z>>\lambda_F$ and with $a$ being a system-dependent constant, thus confirming the analytical calculations reported in Ref.~[\onlinecite{HPP10}]. This result is in sharp contrast with the {\it universal} (metal independent) asymptotics for metal slabs, where
$V_x^{\text{Slab}}(z) \sim - \; e^2 / z$, for $z \gg d$. An important difference between
these two cases is that whereas for metal slabs the asymptotics are dominated by the
so-called Slater potential $V_x^{\text{S}}(z)$ entering Eq.~(\ref{KLI0}), for the SI metal the asymptotics are dominated by the two remaining contributions $V_x^{\Delta}(z)$ and $V_x^{\text{Shift}}(z)$ entering Eqs.~ (\ref{vexchange}) and (\ref{KLI0}).

As in the case of the exact-exchange energy density~\cite{HCPP09}, we attribute the qualitatively different behavior of $V_x(z\to\infty)$ and $V_x^{\text{Slab}}(z\to\infty)$ to the fact that these asymptotes are approached in different ranges. Although in the case of the SI metal the asymptote is reached at distances $z$ from the surface that are large compared to the Fermi wavelength $\lambda_F$ (the only existing length scale in this model), for metal slabs the asymptote is reached at distances $z$ from the surface that are large compared to the slab thickness $d$. For thick slabs with $d>>\lambda_F$, $V_x^{\text{Slab}}(z)$ first coincides with $V_x(z)$ [dictated by Eq.~(\ref{asint.tot}) at $z>>\lambda_F$] in the vacuum region near the surface, but at distances from the surface that are considerably larger than $d$, $V_x^{\text{Slab}}(z)$ turns to the slab image-like behavior of the form of Eq.~(\ref{slab}).
For increasingly wide slabs, the SI regime extends to larger values of $z$
(in the intermediate range $\lambda_F < z \ll d$, if feasible)
, and finally for $d \rightarrow \infty$, the slab regime is never reached.

Our results and conclusions are in contrast with the claim of Ref.~[\onlinecite{Qian12}] that
for the SI metal the asymptotics of the KS exchange potential $V_x(z)$ are embodied by half the Slater potential $V_x^{\text{S}}(z)$. Our OEP numerical calculations clearly show that this is not the case. Indeed, the SI-metal actual asymptotics are dominated by $V_x^{\Delta}(z)$ and $V_x^{\text{Shift}}(z)$, which after scaling and in the asymptotic region are found to be well fitted by Eq.~(\ref{fit}). In the case of $V_x^{\Delta}(z)$, the coefficients $a^{\Delta}$ and $b^{\Delta}$ entering Eq.~(\ref{fit}) are {\it universal}, i.e., do not depend on the electron density. In the case of
$V_x^{\text{Shift}}(z)$, however, $a^{\text{Shift}}$ and $b^{\text{Shift}}$ they both depend on the electron density even after scaling. For the metal electron densities under study ($r_s=2.07$ and $r_s=6.00)$, there is a net logarithmic contribution to
Eq.~(\ref{fit}) and the leading asymptotic contribution to $V_x(z)$ happens to be of the form
of Eq.~(\ref{asint.tot}).
Based on our experience with the OEP formalism, it should be feasible to perform {\it ab-initio} calculations, such as the ones
presented here but for a real SI metal-vacuum surface, relaxing the jellium approximation. It would then be interesting to check whether  the
$e^2 \ln (az) /z $ asymptotic scaling of the exchange potential found in Ref.[\onlinecite{HPP10}] and confirmed here survives to such strong test, as it happens in the slab case. 

Now that we have a clear understanding of the asymptotics of the KS exchange potential of DFT
for both metal slabs and a SI metal, the next ambitious step is to investigate the contribution to the asymptotics coming from correlation. In the case of three-dimensional finite systems, correlation is known not to contribute to the leading asymptotics. 
This consideration is not applicable, in principle, neither to the slab which is finite 
along $z$ but extended in the perpendicular plane nor to the SI metal-vacuum case, which is 
extended in the three spatial directions. Whether in the case of a SI metal surface correlation brings, in the asymptotic region, $V_{xc}(z)$ to the classical image potential of the form $-e^2/4z$ we do not know yet. Work in this direction is now in progress.

\section{Acknowledgments}

C.M.H. wishes to acknowledge financial support received from CONICET of Argentina through PIP 2014-47029.
C.R.P. wishes to acknowledge financial support received from CONICET and ANPCyT of Argentina through Grants No.PIP 2014-47029 and PICT 2016-1087.




\begin{thebibliography}{[Vo]}
\bibitem{note1} Note that this Hartree potential includes the contribution coming from the uniform positive background (proportional to $n_{+}$). 
Alternatively, this contribution could have been  denoted separately as the "external potential".
\bibitem{note4} Mean values are defined as $\overline{O}^{\, i} = \int \xi_i^{*}(z) \, O \, \xi_i(z) \, dz$.
\bibitem{note3} Deep in the bulk, the electronic density goes to its bulk value, $n(z \rightarrow -\infty) \rightarrow \overline{n}$. 
From Eq.~(\ref{de}), this implies that ${\Delta V}_x^k(z \rightarrow -\infty) \rightarrow \overline{\Delta V}_x^k$, which in turn implies that $\Psi_k(z \rightarrow -\infty) \rightarrow 0$.
\bibitem{note2} $r_s$ is a useful dimensionless parameter, defined as $r_s := r_0 / a_0$, with $r_0$ being the radius of a sphere containing on average one electron.
It is related to $\overline{n}$ by $r_s = (3/4 \pi a_0^3\overline{n})^{1/3}$
and to the Fermi wave number by $a_0 k_F =(9 \pi / 4)^{1/3} / r_s$.
\end{thebibliography}
\end{document}